\documentclass[authorversion, screen]{acmart}

\usepackage{acronym}
\usepackage{amsmath}
\usepackage{mathtools}
\usepackage{tabularx}
\usepackage{acronym}
\usepackage{bm}
\usepackage{makecell}
\usepackage{tabularx}
\usepackage{booktabs}
\usepackage{enumitem}

\usepackage{graphicx}

\setlist{nolistsep}
\definecolor{myred}{RGB}{155,0,20}

\newcolumntype{Y}{>{\centering\arraybackslash}X}

\newcommand{\bi}{\begin{itemize}}
\newcommand{\ei}{\end{itemize}}

\newcommand{\Sec}[1]{Sec.~\ref{#1}}
\newcommand{\Fig}[1]{Fig.~\ref{#1}}

\newcommand{\scaledwidth}[1]{#1\columnwidth}

\acrodef{ACS}{ant colony system}
\acrodef{VM}{Virtual Machine}
\acrodef{AC}{Application Component}

\AtBeginDocument{%
  \providecommand\BibTeX{{%
    \normalfont B\kern-0.5em{\scshape i\kern-0.25em b}\kern-0.8em\TeX}}}

\setcopyright{acmcopyright}
\copyrightyear{2022}
\acmYear{2022}
\acmDOI{XXXXXXX.XXXXXXX}

\acmPrice{15.00}
\acmISBN{978-1-4503-XXXX-X/18/06}

\begin{document}


\title[\mbox{Energy-Efficient} Mobile Edge Computing: State of the Art and Open Challenges]{Energy Efficient Deployment and Orchestration of Computing Resources at the Network Edge: a Survey on Algorithms, Trends and Open Challenges}

\author{Neda Shalavi}
\authornote{Corresponding authors.}
\email{neda.shalavi@phd.unipd.it}
\orcid{0003-3486-1450}
\author{Giovanni Perin}
\authornotemark[1]
\email{giovanni.perin.2@phd.unipd.it}
\orcid{0002-7333-3004}
\author{Andrea Zanella}
\email{andrea.zanella@unipd.it}
\orcid{0003-3671-5190}
\author{Michele Rossi}
\email{michele.rossi@unipd.it}
\orcid{0003-1121-324X}
\affiliation{%
  \institution{\\Department of Information Engineering, University of Padova}
  \streetaddress{via Gradenigo 6/b}
  \city{Padova}
  \state{PD}
  \country{Italy}
  \postcode{35131}
}








\renewcommand{\shortauthors}{Shalavi, et al.}

\begin{abstract}
  Mobile networks are becoming energy hungry, and this trend is expected to continue due to a surge in communication and computation demand. Multi-access Edge Computing (MEC), a key component of $5G$ and $6G$ networks, will entail energy-consuming services and applications, with non-negligible impact in terms of ecological sustainability. In this paper, we provide a comprehensive review of existing approaches to make edge computing networks greener, including but not limited to the exploitation of renewable energy resources, and context-awareness (e.g., user mobility, or dynamics in the traffic), analyzing their pros and cons. We hence provide an updated account of recent developments on MEC from an energetic sustainability perspective, addressing the initial deployment of computing resources, their dynamic (re)allocation (resource scheduling), as well as distributed and federated learning designs. In doing so, we highlight the energy aspects of these algorithms, advocating the need for energy-sustainable edge computing systems that are aligned with Sustainable Development Goals (SDGs) and the Paris agreement. To the best of our knowledge, this is the first work providing a systematic literature review on the efficient deployment and management of energy harvesting MEC, with special focus on the deployment, provisioning, and scheduling of computing tasks, including federated learning for distributed edge intelligence, toward making edge networks greener and more sustainable. At the end of the paper, open research avenues and challenges are identified for all the surveyed topics.
\end{abstract}


\maketitle

\section{Introduction}
\label{sec:introduction}
The world is at the dawn of a new era. Distinct transformations in the digital technologies we use daily are happening at an exponential pace. New applications and services, such as autonomous driving, Virtual Reality, cyber-physical systems, and Industry 4.0, hold the potential to revolutionize our society in many different directions, improving the quality of life and the efficiency of our production and government processes. Such a revolution heavily relies on Information and Communication Technologies (ICT), which provide the tools to generate, collect, process and exchange data from the most disparate sources efficiently. Hence ICT is demanded to deliver unprecedented performance levels in terms of availability, reliability, low latency, integrity, scalability, safety, and so on. Today, however, these classic performance indicators are no longer sufficient to guide the design of modern ICT systems. Still, they must be complemented by new indicators related to the \textit{environmental sustainability} of the proposed solutions. As a matter of fact, ICT has a non-negligible impact on our plant's ecosystem, accounting for a fraction that is estimated in the range between 1.8\% and 3.9\% of the global Greenhouse Gas (GHG) emissions, according to~\cite{FREITAG2021100340}. As reported in~\cite{Lotfi}, if the Paris agreement is followed (rather optimistic due to current trends), the total worldwide energy footprint in 2040 would remain at the level of 2015. The ICT carbon footprint would amount to $14$\% of the total worldwide footprint in 2016, which is half of the current total footprint of the industrial sector in the United States. Energy efficiency and environmental sustainability should therefore be driving aspects in the design of next-generation ICT systems. But what are the elements that most affect the ICT's environmental footprint? Still, in~\cite{Lotfi}, the authors studied the GHG emissions of different parts of ICT networks from 2010 to 2020, revealing that \textit{communication networks} and \textit{computing facilities} are the main contributors to the environmental footprint of the entire ICT sector, as better explained below.
 
\subsection{Energy cost of mobile communications}
According to~\cite{Lotfi}, communication networks (including Customer Premises Access Equipment, backbone links, office networks, telecom operators, and their cooling devices) are responsible for the second-largest contribution to the total footprint of ICT, going from the 28\% of the ICT-originated total GHG emissions in 2010 to the 24\% of 2020. Operators forecast that, by moving toward fifth-generation (5G) networks, the total energy consumption of the mobile networks will increase by a factor of $150\%-170$\% until $2026$~\cite{170}. Hence, mobile networks consume a considerable amount of energy, and this trend is expected to continue (if not grow) due to a surge in communication and computation demand. It is therefore fundamental to increase mobile networks' energy efficiency to limit their current and future environmental impact.  

\subsection{Energy cost of computing facilities}
If mobile communication is responsible for a significant fraction of the ICT-related energy demand, computing is undoubtedly not secondary in this cheerless race. Still, referring to \cite{Lotfi}, data centers (including servers, communication links between them, storage, and cooling devices) were indeed the main contributors to \mbox{ICT-originated} GHG emissions in 2020, with a percentage that increased from $33$\% to $45$\% from 2010 to 2020. A recent study about the water and carbon footprints of the US data center industry reports that the amount of energy required by Data centers to operate accounts for around 1.8\% of electricity use in the whole nation~\cite{Siddik_2021}. On the other hand, \cite{9744492} reveals that, while the overall data-center energy consumption has constantly been rising over the past years, carbon emissions from operational energy consumption have fallen, mainly thanks to the use of renewable energy. However, as remarked in~\cite{Siddik_2021}, resorting to renewable energy sources may also exacerbate some problems, e.g., local water scarcity and water stress in many watersheds due to increases in water demands to supply the required energy to data centers. Therefore, while exploiting renewable energy sources certainly has a positive effect in reducing the carbon footprint of the ICT sector, the environmental sustainability of such systems can be improved only by adopting a holistic approach that considers the problem's different facets. 

\subsection{The emerging role of Edge Computing}
One trend that cannot be neglected in this scenario is the progressive convergence of mobile communication and computing systems. As a matter of fact, we observe that, although data processing has been traditionally accomplished via cloud servers, recent developments in Multiaccess Edge Computing (MEC) bring computing capabilities to the communications network edge, i.e., closer to the user equipment (UE), where information is generated. MEC servers can be deployed at Base Stations (BSs) or aggregation points between neighboring BSs. The literature on MEC management refers to a number of different scenarios. Still, they generally consider local nodes, edge nodes, and cloud facilities, which have different energy, transmission, data storage, and computational capabilities. Large cloud servers have high energy efficiency per CPU cycle. In addition, data need to traverse across the core, metropolitan, and access networks to move from the end user to the cloud, and this is {\it costly} in terms of delay and energy, as every packet is to be processed and relayed by a large number of network devices (switches, routers, fiber amplifiers, and regenerators). Instead, MEC servers are expected to be less powerful and less efficient in processing high data volumes but are physically deployed close to the access sites where user data is gathered in the first place. Also, while cloud servers are more energy efficient, this does not necessarily mean they consume less. In fact, roughly $30$\% of their energy consumption is attributed to their cooling systems. At the edge, cooling systems are expected to be simpler and, consequently, less energy efficient, but also with lower consumption, which might be supplied more easily by renewable sources. In this new architecture, the boundaries between computing and communication systems are blurred, and the system tends to become a platform where computational, communication, storage, and energy resources can be combined in a flexible manner, offering different performance trade-offs.

The typical use cases that are considered in the literature are the following: (i) Users/applications need to perform some computational tasks with a local utility. To save energy and time, the tasks can be totally or partially delegated to edge or cloud computing facilities, paying a specific cost (still in terms of time and energy) for transferring data, models, and results. (ii) Users/applications cooperate in training a system-level computing model. The computation is distributed among the users, which then exchange control information to merge their local knowledge and converge towards a common model. The general problem, then, consists of managing the different resources (energy, computational power, storage memory, transmission capabilities) at the local, edge, and cloud levels in order to maximize a specific utility function (e.g., number of computational tasks completed in a unit time), given a set of constraints (e.g., on the total amount of consumed energy, energy efficiency per node, etc.). Clearly, the solution depends on the choice of the utility function and constraints, and the problem can quickly become mathematically intractable, thus requiring approximated or empirical solution methodologies. How to make optimal use of the flexibility of this new computing/communication paradigm, particularly considering environmental sustainability aspects, is an open challenge that has attracted considerable attention in recent periods. 

\subsection{Paper focus and novel contribution} 
\label{sec:contributions}

This article provides an updated account of deployment and orchestration algorithms for edge computing resources in modern mobile networks, with focus on their energy consumption aspects, by also discussing the implications related to using Energy Harvesting (EH) hardware. The inclusion of the energy dimension in MEC design/management is, indeed, disruptive, as energy consumption cuts across every single hardware and software part of the system, so to profile the system properly from an energy standpoint, it is necessary to model all the processes and components involved. With this paper, we offer a systematic review of the fundamental aspects that are to be accounted for when planning, deploying and managing computing solutions in an energy efficient manner, discussing leading works in the literature and identifying open avenues. 

Energy efficiency has been considered in a recent work that reviewed the opportunities and open challenges for integrated \mbox{EH-MEC} 5G networks~\cite{ISRAR2021102910}, presenting a broad range of techniques and covering many technological aspects, including radio resource transmission and management techniques. Nonetheless, edge computing and the management of the related computing processes are only briefly discussed. Further, reference~\cite{Jiang2020EnergyAE} provides an account of energy efficient edge computing systems, but focusing on \mbox{energy-aware} hardware design, compilers, Operating Systems (OS), and middleware, with only one final section devoted to edge applications and their offloading. 

The present article departs from previous review papers, as it concentrates on the \textbf{deployment} and \textbf{orchestration} of computing processes, focusing on \textbf{algorithmic aspects} (resource provisioning and scheduling) rather than on hardware, compilers, or operating systems. To the best of our knowledge, this is the first work providing a systematic and comprehensive literature review regarding the efficient deployment and management of \mbox{EH-MEC} scenarios, with special attention to the \textbf{deployment}, \textbf{provisioning}, and \textbf{scheduling} of computing tasks, as well as to \textbf{federated learning} as a means to train distributed edge intelligence.

The energetic footprint of a MEC system results from the many choices that must be made for its design, deployment, and operation. In this article, we endeavor to describe those choices that can shift the balance between the different performance goals and more significantly influence the energy sustainability of the resulting system. We start by describing, in \Sec{sec:architecture}, a general reference scenario for MEC with EH units, which includes as special cases most of the specific scenarios considered in the recent literature. In addition, we describe the most common models used to characterize the energetic aspects of the system, namely the energy production from solar panels, the energy storage features of batteries, and the energy consumption of edge processors. In \Sec{sec:MEC_deployment_strategies}, we discuss how to deploy the energy harvesting and computing hardware considering the (actual and predicted) traffic demand distributions, users' mobility patterns, energy harvesting trends, nodes capabilities, and so on. This goes under the name of {\it resource deployment}. After this, we discuss in \Sec{sec:resource_provisioning} the \emph{resource provisioning}, i.e., the strategies to pre-allocate network resources, such as communication, computation, and data storage, to applications, servers, and network elements. This entails how these functionalities are \mbox{pre-assigned} to different network elements and, in particular, how to allocate/migrate Application Components to/across the various nodes and how to manage data caching in an effective manner. We emphasize that this differs from what we refer to as \emph{resource allocation} that is covered in the \Sec{sec:resource_allocation}, which instead deals with the \emph{re-distribution} of network, data storage, and processing resources at execution time via suitable scheduling strategies. However, as it will emerge from the text, the distinction between resource provisioning and allocation is sometimes blurry. Subsequently, in \Sec{sec:collaborative_learning}, we analyze the use of the MEC network to train learning algorithms in a distributed and collaborative fashion, discussing the advantages and the limits of a platform with heterogeneous capabilities. Finally, we conclude the paper with \Sec{sec:conclusions}, where we discuss some open challenges that emerge from our study of the literature, and that need to be addressed to reduce the environmental impact and carbon footprint of the fifth generation of communication networks (5G) and beyond, leading to greener ICT ecosystems. 

For the reader's convenience, we have grouped the surveyed \mbox{energy-efficient} techniques into four categories, as detailed in Tab.~\ref{tab:table of content}. The remainder of the present paper reflect this categorization.



\begin{table*}[t]
\begin{center}
\begin{tabularx}{\textwidth}{c|Y}
\toprule
\makecell{edge network deployment \\ (Section~\ref{sec:MEC_deployment_strategies})}& MEC network integration methods in terms of geographic available resources, load and mobility patterns, and required interaction between edge servers  \\
\midrule
\makecell{resource provisioning \\(Section~\ref{sec:resource_provisioning})} & service deployment process on the edge servers, VM placement, and data caching\\
\midrule
\makecell{resource allocation \\(Section~\ref{sec:resource_allocation})} & allocation of computing, storage, communications and energy resources among different tasks and servers with energy efficiency considerations\\
\midrule
\makecell{collaborative learning \\(Section~\ref{sec:collaborative_learning})}& context-agnostic and -aware optimization, federated \emph{vs} fully decentralized\\
\bottomrule
\end{tabularx}
\end{center}
\caption{\textbf{Literature classification.} Classification of the energy-efficient works collected in this survey and relative content. The organization of this paper is structured according to this grouping.}
\label{tab:table of content}
\end{table*}

\section{Reference scenario and Energy Model}
\label{sec:architecture}

In this section, we present general reference scenarios for the study of sustainable edge computing and discuss the underlying assumptions. Then, we briefly overview the state-of-the-art energetic models of the main system components, namely EH (chiefly, solar panels), batteries, and MEC units. We then conclude the section with a summary of the main points regarding sustainable edge computing scenarios. 

\subsection{Reference MEC scenario}
\begin{figure*}[tbp]
\begin{center}
\includegraphics[width=\scaledwidth{1}]{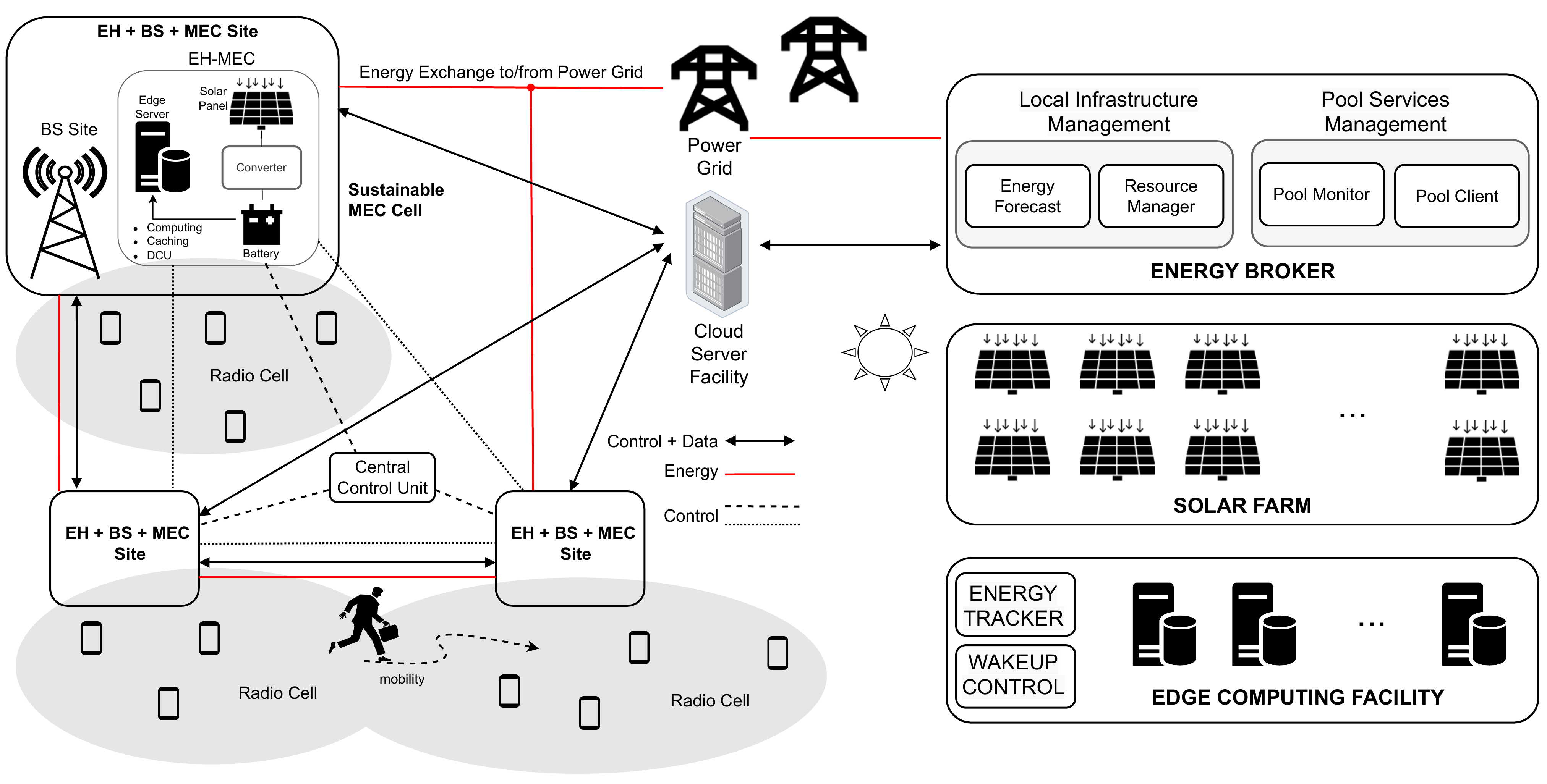}
\end{center}
\caption{\textbf{Left:} a network of Sustainable MEC Cells (SMCs), with BSs equipped with Energy Harvesting (EH) hardware, battery and an edge server (MEC). Edge computing resources can either be managed in a centralized or distributed~\protect\cite{perin2022towards} manner. \textbf{Right:} A novel concept framework for edge computing through distributed renewable energy farms, edge computing facilities and energy broker to intelligently allocate computing resources across edge and cloud facilities~\protect\cite{9352548}.}
\label{fig:network_diagram}
\end{figure*}

The MEC system with Renewable Energy Resources (RERs) depicted in \Fig{fig:network_diagram} entails the different specific scenarios considered in the literature and can hence be regarded as the reference architecture for these types of studies. Below we describe the single elements that compose the system and their role.

\paragraph{Sustainable MEC Cells}
The top-left corner of \Fig{fig:network_diagram} depicts the primary building block of the system, which we refer to as \emph{Sustainable MEC Cell} (SMC). In its basic configuration, an SMC  features (one or more) BSs serving a given group of mobile users. The BSs are connected to an on-site edge server (MEC), which enables local operations such as processing, data caching, and local decision support. Some (or all) BSs are equipped with energy harvesting hardware (solar panels in the figure) that provides RERs for use within the SMC. Surplus energy may also be stored in local energy storage (batteries) for later use. 
Different SMCs may act in a coordinated fashion by sharing their computing resources, selling surplus energy to the power grid, or even transferring energy between sites. In the latter case, a micro-grid may interconnect the BSs so that they can share their energy surplus to reduce the network power intake from the power grid. Energy management algorithms at different levels are required to operate this network to provide communication and computing services to the end users while using the harvested energy efficiently. Such coordination can be obtained through a Central Control Unit (CCU) that orchestrates energy and scheduling decisions across BS sites or via the Distributed Control Units (DCUs) of the SMCs, which cooperate to reach scheduling decisions in a distributed manner. 
A reasonable goal for the scheduler is to allocate as much load as possible to those servers that experience an energy surplus. However, their hardware may not necessarily be the best fit for the computing task. Objective functions should then encode this principle, as well as other criteria like completing the computing tasks within a given deadline, achieving a sparse allocation, i.e., using as few MEC servers as possible to allow the remaining ones to enter a \mbox{power-saving} mode. This approach is adopted, for example, by the authors of~\cite{perin2022towards} who propose a distributed optimization framework to (i) reduce the energy drained by edge computing facilities from the power grid and (ii) distribute the user-generated computing load across edge servers while meeting computing deadlines, and achieving goals like load balancing vs. consolidation in the allocation of computing processes. Edge servers are co-located with the BSs, RERs are available to power BSs and servers, and end users generate workload having processing deadlines. A predictive, online, and distributed scheduler for \mbox{user-generated} computing tasks are devised. It achieves fast convergence, using computing resources efficiently, and obtains (best case) a reduction of $50$\% in the amount of renewable energy that is sold to the power grid by heuristic policies. This surplus of renewable energy is instead used at the network edge for processing to take maximum advantage of RERs.

\paragraph{Photovoltaic-powered distributed MEC plants} 
The right-hand side of \Fig{fig:network_diagram} exemplifies the solution proposed in~\cite{9352548}, where a new approach is devised to reduce the carbon emissions of cloud/edge computing through the deployment of small-scale photovoltaic (PV) plants. The authors advocate moving some cloud services to distributed PV-powered computing facilities. Leasing computing resources is compared to selling surplus energy from the PV plants to the power grid. The {economic and technical viability} of the approach is assessed with a positive outcome.
In detail, a hardware unit, termed local unit (LU), is utilized to monitor the state information of local facilities in terms of PV power production and consumption, on/off state, usage of computing units, as well as to trigger a \mbox{wake-on-lan} control signal. The LU reports this state information to an online and cloud-hosted \mbox{micro-service} called Energy Broker (EB), which (i) estimates the total PV production (kWh/year) using historical data and (ii) based on such estimates, determines the annual payback and the system’s feasibility. Moreover, the EB estimates one-hour-ahead PV production using short-term weather forecasts and current weather observations to determine which computing units will have surplus PV power (e.g., full battery) to join the computing pool. The proposed strategy adds computing units to the pool only if the estimated revenue is higher than selling the surplus energy to the power grid, which is assessed using online grid energy prices. Numerical results indicate that Support Vector Machine Regression (SVR) obtains the highest forecast accuracy among other simple regression approaches. Still, even a multivariate linear regression model achieves similar accuracy with a training phase. In addition, even with a partial allocation of computing units to the pool (in case of low computing demands), the revenue obtained by solar-powered computing overcomes the income accrued by directly selling the green energy surplus to the power grid.

\subsection{System Energy Model}
\label{sec:Emodel}

In order to achieve environmental sustainability, the system has to be designed and managed by adopting a holistic approach, which accounts for all the aspects that impact the generation and consumption of energy. It is hence essential to properly characterize the energetic profile of the main system components. In the following, we recall the energetic models commonly used in the literature for energy production, storage, and consumption. 

\paragraph{EH models}
RERs may be generated in a variety of ways from different renewable sources, such as sun, wind, tides, waterfalls, and so on. The RERs can then be consumed directly at the place where they are generated or transferred to other locations through the power grid. In the MEC literature, RER is generally assumed to be produced by means of solar panels, which can either be distributed among the SMCs (as for the first scenario described in this section) or concentrated in PV plants (as for the second scenario). Nowadays, solar panels can be appropriately modeled (e.g., \cite{solarstat-2014}), and solar irradiation datasets are widely available from several sources, either freely or at little cost as, e.g., in the ``uMASS Data Set for Sustainability''~\cite{uMASS-solar} or the ``NSRDB: National Solar Radiation Database''~\cite{NSRDB}. Hourly energy prices for RERs and the power distribution network are also available, and communication models for BSs are well developed. In principle, therefore, it is possible to characterize quite accurately the energy production of solar plants. However, this may be cumbersome, depending on the specific technology of the solar panels, their size, orientation, and geographical location, not to mention the weather conditions. In many studies, therefore, the authors prefer to resort to tabular traces that can be found in the literature, or abstract stochastic models of the renewable energy sources, which generate energy units according to given stochastic processes (e.g., \mbox{non-homogeneous} Poisson processes). 
A further discussion on available RERs is offered in~\cite{Emodels}, where the authors present different RER types and the corresponding energy models for designing energy schedulers in communication systems and provide guidelines for interested researchers to choose suitable energy models for their use cases.

\paragraph{Energy accumulator models} Although battery models are well known; often in the EH-MEC literature, batteries are assumed to have an ideal behavior. That is, being $E_0$ the current battery level and $\Delta E$ the amount of energy harvested in the last reference period, the new level is set to $E_1 = \min (E_0 + \Delta E , E_{\max})$, where $E_{\max}$ is the maximum battery capacity. So, non-linearity, memory effects in the amount of charge, and its leakage are often not considered. However, they may have a non-negligible impact on optimizing the energy management strategies (see, e.g.,~\cite{6782280}).

\paragraph{MEC energy consumption models}
MEC nodes are assumed to be equipped with computing (CPU or GPU), storage, and communication hardware, whose capabilities are usually assumed to vary, depending on the type of MEC device considered. In this study, we are particularly concerned about the energetic profiling of edge processors. To be noted, most of the research works consider relatively simple energy consumption models for CPU and GPU. For example, in~\cite{9403911} the energy consumption for computing is assumed to be proportional to the percentage of CPU resources that are allotted, i.e., to the number of CPU cycles required to perform a task, a model taken from~\cite{Vogeleer-2013}. Reference~\cite{Dayarathna-2015} provides a full account of such CPU frequency-based model, where the processor (CPU) power is modeled as $P = C V^2 f$, where $C$ represents the switched capacitor, $V$ the processor voltage, and $f$ the running frequency. In~\cite{RiskES}, the processing energy is calculated by adding two contributions: 1) a {\it static} energy component representing messaging, idling, etc., 2) a {\it dynamic} energy component that is modulated by the traffic intensity, being multiplied by CPU coefficients to convert traffic load into energy consumption. Paper~\cite{perin2022towards} also assumes that the energy consumed for processing depends linearly on the workload, which descends from, e.g.,~\cite{Bertran-2013}. We remark that these models are taken from research published in 2013, which refers to cloud servers or mobile systems and do not yet account for the distinctive traits of modern edge computing architectures oriented towards Machine Learning (ML)~\cite{Aria&Ahmed}, as we will discuss in \Sec{sec:conclusions}.

\subsection{Remarks on MEC scenarios}\label{sec:renewables_open_challenges}
Recent research has shown that energy aware techniques to manage MEC resources can bring significant benefits to achieving energy efficiency goals. In particular, the energy and carbon neutrality of edge computing services and MEC facilities seem to be attainable in some practical scenarios. Along these lines, the authors of~\cite{perin2021ease} estimate that carbon neutrality is achievable for hybrid energy harvesting and grid-connected MEC networks over a wide range of their case studies thanks to dedicated hardware and software components. There are different ways to pursue environmental sustainability. From the findings in~\cite{9352548}, one approach could be maximizing the local utilization of RERs, even at the cost of lower energetic efficiency of the whole system. From an environmental perspective, the underlying assumption is that transferring RER surplus to the power grid is less convenient than (even sub-optimal) using RERs where they are produced. However, the design of efficient solutions is complicated by the vast heterogeneity of device capacity and energy efficiency, users demands, and RER productions. In the following sections, we will analyze the approaches proposed in the literature to address this challenging task.

\section{MEC Network Deployment}
\label{sec:MEC_deployment_strategies}

Planning the geographical deployment of a MEC system is a critical task that can strongly impact the system's actual performance and environmental sustainability. The worthiness of a particular MEC deployment can be assessed by means of the performance metrics defined by the European Telecommunications Standards Institute (ETSI)~\cite{etsi6v1}. These metrics can be divided into two groups, namely \textit{Functional metrics}, which assess the \textit{user performance} and include some classical indexes such as latency in task execution, device energy efficiency, bit-rate, loss rate, jitter, Quality of Service (QoS), etc.; and \emph{Non-functional metrics} that, instead, focus on the \textit{MEC network deployment and management}. Non-functional metrics include the following indexes.

\bi
\item \emph{Service life-cycle} considers all the steps of service deployment at the MEC network, including the service boot-time, i.e., the sum of instantiating, deployment, and provisioning time, and the time required to deploy the service and update its resources, when required. 
\item \textit{Resource consumption footprint} accounts for processing units, data exchange among edge hosts, and hosts with wireless nodes or orchestration control units.
\item \textit{MEC energy efficiency} is gauged as the MEC network’s energy consumption per data volume for a particular traffic demand under specific QoS constraints. 
\item \textit{Availability} and \textit{reliability} refer to the probability of finding available computation capabilities and completing processing tasks in due time once the MEC network accepts the assignment, respectively.
\item \textit{Processing/computational load} refers to the distribution of the workload across MEC nodes.
\ei

These ETSI metrics provide a means to quantitatively compare MEC and \mbox{non-MEC} solutions or MEC host placement methods with different optimization objectives in terms of deployment and latency costs, workload allocation, and maximization of economic benefit, etc. Moreover, these metrics indirectly indicate what aspects must be accounted for when planning a MEC deployment. For example, to minimize the resource consumption footprint metric, the number of deployed MEC units should be minimized, and each unit should be massively used to fully exploit its potential. However, workload and network density vary across metropolitan areas. As such, MEC servers might work in an idle state in sparse areas, wasting energy, while heavy load might occur in denser areas, overloading the edge servers. Therefore, to improve the MEC energy efficiency metric, it is advisable to use edge servers with different capacities based on the geographical distribution of the workload, which depends on the users' mobility and demand pattern.
Moreover, the availability of RER sources should also be accounted for to increase the availability of green energy and promote system self-sustainability. From this quick analysis, it is evident that the placement of the MEC network elements should be decided based on a {\it full knowledge} of the network state, which includes traffic loads, users mobility, and environmental measurements, such as weather conditions and harvesting energy profile. Optimizing the MEC deployment is an offline task performed before establishing the MEC site. 

In the remaining of this section, we review the approaches that aim to minimize the idling time of the MEC servers by either considering the geographical distribution of the workload or the users' mobility pattern. We then look into the problem of MEC placement in cell-free massive Multi-Input Multi-Output (MIMO) systems, where the users are not connected to a single BS but are actually served by a subset of all the BSs in its coverage range. Finally, we conclude the section by remarking on the take-home messages that can be drawn from the literature addressed.

\subsection{Geographic Workload Distribution} 
Under-utilized MEC servers spend long periods in an idle state, wasting energy. To minimize this type of energy loss, there are two main possibilities: either reduce the number of MEC servers so as to increase the utilization of those that are deployed; or deploy multiple MEC servers but with different capabilities, matching the workload distribution so that each server is used close to its maximum capacity. Reference~\cite{Li-conf2018} pursues the first strategy by considering the maximization of the edge servers utilization rate as the optimization objective in the server placement. Specifically, an attempt is made to reduce the total energy consumption of MEC networks by allocating multiple BSs to each MEC server so as to maximize its utilization rate while keeping the serving latency lower than a predefined threshold. The authors of~\cite{clusterbasedMECdeployment} focus on a vehicular network where RoadSide Units (RSUs) are equipped with MEC capabilities and solve the server placement problem by jointly considering different optimization objectives. A clustering algorithm for RSUs is proposed to estimate the number of required edge servers. Each cluster has a single edge server acting as its cluster head, and the RSUs with the highest workload within each cluster are selected as candidate nodes for the placement of an edge server.

Furthermore, the cluster radius is set to be inversely proportional to the RSU densities in each region to achieve workload balance with respect to geographical workload distribution. This approach could save unnecessary edge server energy consumption in idle or low utilization conditions. After defining the clusters and the candidate MEC servers positions, the deployment strategy is obtained via a multi-objective algorithm based on adaptive differential evolution. The optimization objectives are edge server's energy consumption, workload balance, processing delay, network reliability, and deployment cost. The edge servers energy consumption is modeled by linearly scaling the consumption from idle to maximum load state, using the CPU utilization rate as the load indicator. Although the proposed technique affects the total network energy consumption, they only provided numerical results on the convergence speed and the geographical distribution of the resources with the proposed clustering algorithm. In \cite{2018conf}, instead, the authors propose a comparative study of different classes of edge servers. First, they compare the total energy consumption of three edge servers with small, medium, and large resource capacity for a variable amount of load running on them simultaneously. From their results, it is deemed reasonable to deploy large capacity servers when a minimum of $340$ Virtual Machines (VMs) run on them. These servers have a low energy consumption per CPU cycle at high load and higher energy consumption with respect to a less powerful server in their idle state. Medium-capacity servers are deemed efficient for running from $143$ to $340$~VMs, and small capacity ones for fewer than $143$~VMs. Second, they compared deploying three small-capacity edge servers against one medium-capacity server in a radio cell with high-variance load demand. In the first case, all three small servers worked at full utilization rate only at peak hours, while two were in sleep mode during low-demand hours. Instead, the medium-capacity server had to stay active for longer to serve even a low demand. The result showed that deploying multiple smaller servers or a mix of different classes of edge servers in radio cells with highly variable load demand turns out to be more energy efficient than deploying \mbox{high-capacity} servers. The method proposed in~\cite{2018conf} reduces the energy consumption of the MEC network by up to $34.32$\% through a careful deployment of MEC servers with heterogeneous capacity.

\subsection{User Mobility Patterns} 
 
Another aspect to consider at edge server deployment time is the {pattern of users' mobility}, especially in urban scenarios. The work in~\cite{mobilityawareMECdeployment} is based on the intuition that urban dynamics of human movement (namely, user mobility patterns) determine the dynamics of the load pattern within the edge network. The authors of this paper investigate the Edge Data Center (EDC) deployment problem in intelligent cities, devising energy efficient and mobility aware strategies. They choose to collocate EDCs with existing BS sites to reuse the cellular infrastructure as much as possible.

They propose a heuristic technique called Energy efficient Mobility Aware Deployment Algorithm (EMDA) and compare it with two other proposed heuristic baselines named: 1) Mobility Aware Deployment Algorithm (MDA) and 2) Distributed Deployment Algorithm (DDA). An {\it outage} is declared if the MEC network fails to fulfill some demand within some latency constraints, e.g., due to the lack of available resources. The objective of the proposed schemes is to minimize the outage probability with constraints on energy efficiency. DDA uses the k-medoid method to cluster BSs and collocates an EDC with a BS within each cluster. However, this method solely relies on the distance among BSs and EDCs, neglecting the load distribution. MDA ameliorates this aspect by computing a cost based on the number of computation requests received by the BSs during the day. Their main proposal, EMDA, allows for saving energy by selectively powering off those BSs that have experienced low traffic in a previous time window and re-routing their traffic to other neighboring BSs that remain active. Specific models are accounted for mobility, computational demand, and EDC energy consumption.

{User mobility} is convincingly modeled as a graph whose vertices represent points of interest. Different mobility behaviors are accounted for: random walking, uniformly distributed patterns, or real mobility traces. The {computational demand} is gauged based on the aggregated traffic from user tasks. Each EDC serves the computational demand generated by the users connected to the set of BSs that it is responsible for. User activities and their social interactions are characterized using synthetic data traces available in the literature for augmented reality, health applications, compute-intensive popular applications, and actual data traces for a video streaming application. Based on these, the hourly load inside their considered geographic area is estimated. However, little details are provided on the way the task intensities are modeled, i.e., the amount of workload that the tasks entail for the server and the time needed to execute them. The power required by EDCs is modeled as:

\begin{equation}
  P_{\textrm{server}}=P_{\textrm{idle}}+\frac{P_{\textrm{peak}}-P_{\textrm{idle}}}{2} (1+\lambda-e^{-\lambda/\tau}),
  \label{eq:server_power}
\end{equation}
where $\lambda$ is the server's load and $\tau$ is a constant typically in the range $[0.5,0.8]$. Eq.~\eqref{eq:server_power} is representative of the power required by a server using dynamic voltage and frequency scaling. 
Their simulation results indicate that the DDA algorithm leads to the highest outage probability due to neglecting workload balance among EDCs. The MDA technique reduces outages over DDA,  and EMDA achieves better energy savings by turning off BSs experiencing low load. With this approach, some BSs consume $15.7$\% more energy, while the overall daily energy consumption reduces by $27$\%. The results are obtained via a purposely developed simulator called CrowdEdgeSim, which jointly models urban mobility, network, and computing processes. 
\subsection{Cell-free MEC Networks} Unlike the previously described works, edge server deployments in cell-free massive MIMO networks entail a different usage model. In these setups, end users have access to all Access Points (APs) in their coverage range and can communicate with them concurrently. Also, they can either offload a portion of their computing tasks to multiple nearby edge servers or entirely offload tasks to a single edge server. The authors of~\cite{cellfree} studied the impact of AP's coverage range on MEC network energy consumption and the probability of successful task completion at the edge. They compared two scenarios. In the former, APs are equipped with many antennas to provide extensive coverage ranges, leading to {\it sparse} MEC server deployments. In the latter, APs have fewer antennas, leading to shorter coverage ranges and, in turn, to {\it high-density} MEC server deployments. Numerical results demonstrate that high-density deployments consume less energy for any desired successful task computing probability. That is, the total energy cost to successfully compute the tasks in the sparse deployment scenario is much larger than that required to keep the edge servers online in a high-density setup.


\subsection{Remarks on MEC Deployment}
\label{sec:MEC_deployment_open_challenges}

MEC deployment revolves around looking for the best possible locations where to place edge servers to satisfy the QoS/low latency requirements of end users. Most of the works in the literature consider service QoS, load balance, response delay~\cite{satellite}, cost efficiency~\cite{costE}, installation cost~\cite{installationcost}, and maximization of overall profit~\cite{profit}. From the literature reviewed, we learned that the capacities of MEC servers should match the workload demand in order to maximize the overall energy efficiency of the network by reducing the time and energy spent by servers in an idle state. In addition, rather than a few high-capacity MEC servers, it may be more efficient to deploy multiple low/medium-capacity servers, which would allow a finer distribution of workload, maximizing energy efficiency. This result is also confirmed in the case of cell-free massive MIMO systems, where a dense deployment of MEC servers is apparently more efficient than a sparse one. Knowledge of user mobility patterns is another resource that can be leveraged to determine where to allocate edge data centers and how to selectively shut down BS and MEC units not needed to handle user demand. A summary of the impact of MEC deployment techniques is given in Tab.~\ref{tab:impact_of_MEC_placement_techniques}. We observe that these studies generally do not consider the actual environmental cost of device production, operation, maintenance, and decommissioning, which can have an important impact on the overall carbon footprint of the network.

\begin{table*}[t]
\footnotesize
\begin{center}
\begin{tabularx}{\scaledwidth{0.85}}{Y|c|c|c}
\toprule
\textbf{Papers}&\textbf{Objective}&\textbf{Metric} & \textbf{Impact} \\
\midrule
~\cite{Li-conf2018}& utilization rate & \makecell{distance (latency), energy,\\ workload}  & \makecell{single MEC server serving multiple BSs,\\reducing idling periods(energy waste)}\\
\midrule
\cite{clusterbasedMECdeployment}& \makecell{clustering+ multi-\\objective placement} & \makecell{delay, energy, load density,\\ reliability, monetary cost} &  \makecell{increased energy efficiency\\ reducing idling periods }\\
\midrule
\cite{2018conf}& \makecell{QoS/energy tradeoff} & \makecell{delay, mobility pattern,\\energy, processing power} & \makecell{increased energy efficiency,\\ reducing idling periods}\\
\midrule
\cite{mobilityawareMECdeployment}& \makecell{clustering+ minimizing\\service outage} & \makecell{delay, energy, mobility pattern} &  \makecell{increased energy efficiency\\ reducing idling periods }\\
\midrule
\cite{cellfree}&task completion/energy trade off&\makecell{delay, energy}& \makecell{decreased communication energy costs}\\
\bottomrule
\end{tabularx}
\end{center}
\caption{Impact of proposed MEC deployment strategies from the literature on energy consumption.}
\label{tab:impact_of_MEC_placement_techniques}
\end{table*}

\section{Resource Provisioning}
\label{sec:resource_provisioning}

 The second step after deploying edge servers across the network is distributing cloud services/applications replicas among them. Edge servers have a certain amount of computation, communication, storage, and RERs. Distributing various replicas of services across the MEC network affects resource utilization and performance metrics. The popularity of services and requested data among users with geographically varying load demands provides insights on {\it where to place} service replicas at the network edge. Careful service placement with respect to available resources at the edge, i.e., {\it resource provisioning}, enables efficient communication between users, edge and cloud, and efficient computing and storage resource utilization. In this section, we discuss algorithms that primarily focus on efficient resource provisioning at the edge.

\subsection{Service Deployment}
\label{sec:application_components}

Servers allocate a share of their computing resources by dedicating Virtual Machines (VMs) or containers to each user application, as these virtual environments enable dynamic scaling of computing resources based on the user demand~\cite{IEEEsurvey}.   A single software service may involve several ACs, which may be run in parallel, in sequence, in loops, etc., and usually executed inside a VM. Efficient placement algorithms are required to install ACs onto the infrastructure nodes across an end-to-end cloud-edge network. The \textit{fog computing} paradigm tackles the mentioned computation/transmission energy tradeoff by deploying some ACs at the network edge on \mbox{so-called} ``fog nodes'' while running other ACs on cloud servers. We remark that moving ACs from cloud facilities to edge servers might impact the energy consumption of different network devices based on the cost of data forwarding from access to core, the different efficiencies of edge and cloud servers, including their cooling systems, RERs availability, and so on. These aspects are better discussed below.

\paragraph{Cloud \textit{vs} edge VM placement} 
The authors of~\cite{archref} explore this trade-off by developing a framework based on mathematical modeling and heuristics to assess whether and how to place VM services (cloud \textit{vs} edge) with the objective of {\it minimizing the overall energy consumption of service provisioning}. In their framework, the term ``placement'' refers to the process of migrating VMs from cloud to edge and replicating VMs into multiple copies at the edge. They characterize the energy consumption of all network components and study the effect of application popularity, workload profile, and required transmission data rate on the energy-efficient VM/AC placement. The migration of VMs from cloud to edge servers is actually obtained by placing VM replicas on the edge nodes. This reduces the network energy consumption since only the state of the VM needs to be transferred to the clone at the edge but potentially increases the processing energy due to the multiple VM replicas, leading to a trade-off. The network energy consumption is split into (i) {\it traffic-induced} and (ii) {\it processing-induced} consumption. Hence, an offline energy-efficient AC placement algorithm based on Mixed Integer Linear Programming (MILP) with constraints is proposed to classify ACs according to their popularity, minimum workload requirement, and data rate. They consider CPU workload profiles that are either constant or linear with the number of active users. ACs with linear profiles have a baseline power requirement related to their minimum CPU usage (workload baseline) when idling. Then their CPU workload increases linearly with the number of users. Workload baselines as a percentage of the server CPU capacity are estimated to be $1$\% for database applications, $5$\% for website applications, and $40$\% for video applications. The results of the proposed placement scheme indicate that creating multiple replicas at the edge for ACs with a $1$\% workload baseline is energy efficient as this amounts to a negligible overhead for each replica.

Further, the number of AC replicas placed on edge servers should decrease as their workload baseline increases. That is, creating multiple replicas for ACs with high, medium, and low popularity/data rate transmission requirements on edge servers, metropolitan, and cloud servers is respectively found to be the best solution in terms of energy efficiency. Also, there is a trade-off between reducing traffic-induced energy consumption, e.g., moving the data to the edge, and the increase in processing-induced energy consumption due to placing multiple replicas. To conclude, a real-time and \mbox{low-complexity} AC placement algorithm is proposed, which matches each AC with a class of ACs from the offline optimization step and places it based on the class placement strategy. This algorithm saves up to $64$\% of the network energy consumption of \mbox{cloud-only} AC placement.

\paragraph{ACs relocation for load balancing} The authors of~\cite{DECA} study the relocation of ACs, which may be required to achieve a more balanced allocation in response to time-varying load. This is accomplished by (frequently) exchanging traffic matrices containing the required data by the ACs being executed. They propose an optimization algorithm for the initial placement and migration of ACs among servers. The objective is {to minimize the energy consumption and carbon emission rate} of the servers and their communication links with regard to the initial placement and relocation of ACs. ACs are allocated based on the amount of resources they require, their traffic matrices (plus network state), energy prices, and carbon footprint rate per energy unit at geographically distributed edge servers. The proposed algorithm reallocates ACs running on overloaded or underutilized servers by placing (new allocations) / migrating (existing allocations) ACs with intensive communication requirements on/to adjacent edge servers as this reduces the \mbox{inter-AC} communication burden while also using already active network switches. Through the adoption of these strategies, servers and communication links switch to active mode in a greener fashion. Taking the energy budget of computing nodes into account is an effective way of utilizing the green energy harvested by MEC nodes or managing the limited energy budget supplied by their batteries, avoiding an excessive energy consumption from the power grid. The authors of~\cite{Liu-MADRL-ACP} propose AC migration algorithms to minimize users' task completion time subject to the migration energy budget. They present a low complexity multi-agent policy gradient optimization algorithm based on actor-critic reinforcement learning. Distributed agents, referred to as ``actors'', learn how their migration decisions contribute to the global goal of minimizing all the task completion times in a personalized and distributed fashion. A centralized critic computes the effect of agents personalized decisions on the global goal and feeds it back to the agents to steer their actions based on it.

Moreover, a shared \mbox{actor-network} is used for all agents to accelerate the learning process. This proposal improves the task completion time by up to $30-50$\% with respect to other reference methods having a constraint on the energy budget. The reason for their successful result lies in taking advantage of cooperation among MEC nodes to achieve the global goal. In particular, the proposed algorithm significantly reduces the task completion time as the energy budget increases. The proposed approach could be extended to solar EH-MEC networks by additionally accounting for the energy income from RERs.

\paragraph{Energy-aware and mobility aware AC placement} Another work~\cite{stochasticEAappplacement} modeled energy-efficient and mobility-aware AC placement and migration as a multi-stage stochastic programming problem. To capture the distribution of user mobility and their demand from MEC servers, recorded data of taxis in San Francisco was used as a case study. MEC servers in the considered network are heterogeneous in their energy budget and computation capacity. The optimization parameters in each time slot are known, and the uncertain parameters in future time slots are estimated using user mobility distribution and AC relocation costs of the recorded data. The objective function is designed to make placement and relocation decisions by maximizing the sum of individual QoS perceived by the users in the current time step and minimizing the expected AC relocation cost in future steps. The computation capacity of edge servers is modulated according to their available energy budget, which is considered a constraint for optimization. A parallel greedy sample average approximation is exploited to solve the optimization problem: $H$ samples are generated, each including $L$ independent scenarios with different AC locations for each user at each time slot. A graph is then constructed to present the relocation cost among edge servers in the considered time slot for each scenario. Hence, these generated scenarios are used to solve an equivalent deterministic problem by determining the value of the objective function in each case. The scenario achieving the best result is selected as the candidate solution. 

A similar study (\cite{2018conf}) proposes a joint three-step AC and server management algorithm  {\it to minimize the total energy consumption of the MEC network}, which consists of (i) AC migration queuing, (ii) AC placement and (iii) server activation, without significantly degrading the user QoS. If a server is overloaded, the best fit AC among those running on the server is identified and sent to a migration queue for its relocation to another server. The servers running below an energy-efficient utilization rate are put into sleep mode: the ACs running on these servers are added to a migration queue and then migrated to other servers with sufficient energy resources. New AC arrivals are added to the migration queue as well. The algorithm sorts the list of all active servers and the migration queue based on available and required resources, respectively. Then, ACs are allocated to those servers leading to the smallest energy consumption given the AC utilization rate requirements. In case the total amount of resources required by the ACs on the migration queue is larger than those available on all active servers, a server activation controller activates the sleeping servers with the minimum necessary energy consumption. The management algorithm has a supervisor unit running on one of the servers, which triggers the server activation controller at any new AC arrival, to minimize the AC waiting time (from arrival to being put into service). AC queuing and placement processes are initiated whenever the migration queue changes from empty to occupied (an active trigger control reduces unnecessary resource consumption by the management algorithm). This method saves up to $16.15$\% in total energy consumption during peak hours.

\subsection{Data Placement and Caching}
\label{sec:data_placement}

Caching popular content at the edge servers reduces vertical traffic between the MEC network and cloud facilities, allowing for a decreased latency in serving the user requests. Nevertheless, MEC servers usually have limited storage capacity, and the scarcity of storage space might increase their computing latency for specific tasks. MEC servers might have to increase their operating frequency to compensate for this, resulting in higher power requirements. As a result, a question arises: {\it is it energy efficient to cache content at the edge?} The authors of~\cite{7247608} address this question, proving that caching content at the network edge is indeed efficient. Their work demonstrates that the overall energy required per area in cache-enabled networks is lower than in traditional (non-caching) networks while also allowing for higher area spectral efficiency. Their results also indicate that the energy gap between cache-enabled and traditional servers is more significant for a properly tuned transmit power at the physical layer connecting users to their serving base station. Increasing the size of the caching catalog results in higher energy efficiency, but the slope of this energy efficiency increase reduces at larger catalog sizes. This, in turn, raises the question as to which amount of edge servers’ storage capacity would lead to the best network and energy efficiency performance. To date, and to the best of our knowledge, no clear answer was provided to this. 

Preliminary insights towards answering this question are provided by the comparative results in~\cite{transcoding}. This paper proposes a joint caching and transcoding scheduling heuristic for HTTP adaptive video streaming at the network edge. It seeks to maximize the network energy efficiency by choosing between caching content or not, transcoding it, or retrieving it from a cloud server. Numerical results indicate that it is more energy efficient to cache all the video segments at the edge for small videos. As the videos get larger, transcoding them or recalling video segments from the cloud tends to be more energy-efficient. Finally, the authors of~\cite{cacheEE} analyze the energy consumption of cache-enabled MEC networks. They propose a content caching, wireless, and backhaul links’ transmission scheduling algorithm and compare the energy saved by this method against an optimal policy, random caching, and non-caching. They characterize energy consumption to cache content and deliver cached content through wireless and backhaul links. In case the user content is cached at the MEC servers, a path with minimum energy consumption is selected to deliver it to them. If the content is not cached, it is retrieved from cloud servers. The orchestrator unit is an external entity with full knowledge of wireless/backhaul link maps and global cached content maps. It schedules the communication links and servers to deliver content with minimum energy consumption. Simulation results show \mbox{near-optimal} performance and substantial energy savings against competing algorithms.

\begin{table*}[t]
\footnotesize
\begin{center}
\begin{tabularx}{0.9\textwidth}{Y|c|c|c}
\toprule
\textbf{Papers}&\textbf{Objective}&\textbf{Metric} & \textbf{Impact} \\
\midrule
~\cite{archref} & \makecell{minimizing total network\\energy consumption} & \makecell{energy, VM load profile,\\VM traffic and data rate,} & \makecell{increased energy efficiency, CPU\\and communication link utilization rate}\\
\midrule
\cite{DECA}& \makecell{minimizing energy monetary cost\\and carbon emission rate} & \makecell{energy price, carbon emission rate} & \makecell{reducing carbon footprint and\\inter-AC communication overhead}\\
\midrule
\cite{Liu-MADRL-ACP}& \makecell{delay/energy trade off} & \makecell{delay, energy} &  \makecell{reduced latency, increased energy efficiency}\\
\midrule
\cite{stochasticEAappplacement}& \makecell{minimizing delay with\\constraint on RER budget} & \makecell{delay, energy, mobility pattern} &  \makecell{max RER utilization and\\reducing grid energy consumption}\\
\midrule
\cite{2018conf}& \makecell{minimizing energy consumption\\with delay constraint} &\makecell{delay, energy, CPU capacity}& \makecell{increased energy efficiency\\specially at peak hours}\\
\midrule
~\cite{7247608}& \makecell{Area power consumption\\with QoS constraint} & \makecell{Coverage probability, Energy, QoS} & \makecell{Increased energy efficiency\\and area spectral efficiency}\\
\midrule
\cite{transcoding}& \makecell{Caching/transcoding scheduling\\with energy constraints } & \makecell{ Energy, Caching and\\transmission variables} & \makecell{High energy efficiency with reduced\\cached data and backhaul link usage}\\
\midrule
\cite{cacheEE}& \makecell{Content placement and\\communication link scheduling} & \makecell{Energy, Caching and\\transmission variables} &  \makecell{Optimal solution with\\substantial energy savings}\\
\bottomrule
\end{tabularx}
\end{center}
\caption{Impact of proposed resource provisioning techniques from the literature on energy consumption.}
\label{tab:resource provisioning}
\end{table*}

\subsection{Remarks on resource provisioning}
\label{sec:resource_provisioning_open_challenges}
 
Tables \ref{tab:resource provisioning} summarizes the proposed AC placement and content caching techniques at the edge and their impact on system performance. We observe that studying the regional service popularity, network states, time, and geographically varying load patterns provides valuable insights on where {\it to place service VM replicas} and {\it cache popular} content. Service VM replicas should be reallocated among active and underutilized MEC servers due to the time-varying nature of network state, user mobility, and load pattern. However, migrating VM replicas across servers due to user mobility increases the network energy consumption. Considering hybrid task execution options for different tasks could be a valuable option, leading to a significant reduction in the energy consumption. Authors of~\cite{NOMACHINA} proposed a task scheduling algorithm in a vehicular MEC system, comparing it via simulation against two other scheduling algorithms. In one of the reference methods, edge servers accept and execute all requested tasks in their coverage area without migrating them to nearby servers and transmit the computed results to other edge servers if the user leaves its coverage area. Their results demonstrate that this reference method achieves less energy consumption than others. However, due to load congestion and queuing delays, it performs worse than the other two methods in terms of task completion time. In light of these findings, prioritizing the local execution of tasks (no offloading to other servers) might increase the system performance, leading to higher energy efficiencies.

\section{Resource Allocation}
\label{sec:resource_allocation}

Provisioning resources and services are among the network initialization steps. However, the time-varying nature of network processes, resources, and the offered load require another level of online management techniques to utilize network resources efficiently and dynamically -- such techniques are usually dubbed resource allocation/scheduling algorithms. Although careful resource and service provisioning allows the network to operate within the desired performance boundaries, dynamic resource scheduling algorithms can push the network to operate close to the most efficient operating points at all times, redistributing resources according to time-varying processes. Resource scheduling algorithms may be implemented in a centralized (a single controller acts as the scheduler for a group of MEC-capable BS sites), semi-distributed, or fully distributed fashion. Resource allocation techniques take two main flavors (i) offloading (in full or in part) the end users' tasks to edge servers and (ii) reallocating computing resources and moving tasks across servers. We review task offloading algorithms in section \ref{sec:resource_allocation_static}. The driving principles for these processes are: load balancing,  job queue overflows, lack of processing/storage/energy resources at the edge servers, mobility-related optimizations, and service handover. In section \ref{sec:mobility_aware_scheduling}, we review \mbox{mobility-aware} resource management techniques. Furthermore, in most general cases, resource scheduling algorithms should simultaneously consider the {\it joint} reallocation of all types of resources: these algorithms are discussed in section \ref{sec:joint_optimization}

\subsection{Task Offloading schedulers}
\label{sec:resource_allocation_static}

To reduce the energy consumption of end devices, users' tasks can be partially or totally offloaded to the edge servers. This helps reduce the computation burden on end devices, which usually have limited processing and energy resources while meeting processing deadlines. In some cases, users' smart devices can execute a portion of their tasks. The term {\it offloading ratio} indicates the fraction of a task that is offloaded to the edge server(s) versus the remaining part that is computed locally at the end device.


\paragraph{Centralized schedulers} The authors of~\cite{9197634} propose a constrained \mbox{mixed-integer} \mbox{non-linear} constrained program to jointly set the task offloading ratio, transmission, and computing parameters subject to latency requirements, processing power, and memory availability. The problem is solved via a meta-heuristic algorithm that seeks to decrease the overall energy consumption across end devices and edge servers by leveraging the end users' resources to their maximum extent. It is assumed that tasks can be decomposed into independent parts which can be executed in parallel and partially offloaded. The solution provides good schedules, achieving better energy savings with respect to using other meta-heuristic solvers. The processing energy consumption per CPU cycle is estimated as $\kappa f_{\rm cpu}^2$, where $\kappa$ is a hardware dependent constant and $f_{\rm cpu}$ the CPU frequency. This model may be inadequate to represent modern CPU and GPU architectures.

Moreover, the obtained policies achieve the desired behavior for a small/medium number of end devices. Still, additional methods should be devised to distribute the load, so as to keep the complexity low in a distributed setup with many end nodes. In addition to optimizing the offloading ratio, a wise selection of the server(s), where to offload the task (or portion thereof) can further reduce energy consumption. The authors of~\cite{9380662} propose an \mbox{end-to-end} Deep Reinforcement Learning (DRL) algorithm to maximize the number of executed tasks that are completed on time while minimizing the associated energy consumption. This method assumes monolithic tasks (no partitioning allowed) and leaves the study of tasks that can be partitioned as future work. Thanks to the use of DRL, this algorithm does not require modeling the MEC system dynamics (\mbox{model-free} optimization), which are learned sequentially. The algorithm reduces the energy cost of executed tasks through optimal server selection and computing power allocation for the task offloading. It uses the same energy model of~\cite{9197634}, and may have scalability problems due to the centralized nature of the DRL solver. The authors of~\cite{9054646} propose a centralized intelligent energy management system to control the energy consumption and flow in a two-tier mobile cellular network. A deep reinforcement learning algorithm is devised to maximize the use of the green energy harvested by observing and interacting with the surrounding environment. The algorithm objective amounts to reducing the energy drained from the power grid, subject to traffic level constraints. The environment is initially described through the harvested energy from each BS, the BS battery levels, and the load experienced by the BSs across all hours of the day. The impact of different environment representations on network performance is then investigated. Numerical results indicate that considering the battery level and hour of the day is sufficient to achieve high performance.

\paragraph{Distributed schedulers} In~\cite{perin2022towards}, the authors propose a fully decentralized optimization framework based on \mbox{Douglas-Rachford} Splitting (DRS) to allocate the computing resources of edge servers and decide where to offload tasks subjected to a time deadline. Their dynamic and \mbox{multi-variable} heuristic optimization algorithm distributes workload among MEC servers to achieve load balance or server consolidation while encouraging RER utilization and reducing transmission costs. Their solution employs Model Predictive Control (MPC), showing high performance even with simple predictors, e.g., just knowing the average value of the incoming processing load and energy from renewable resources. The algorithm's low complexity and fast convergent nature allow energy-efficient utilization of EH-MEC facilities. Numerical results show a $50\%$ reduction in the RER energy that is sold to the power grid, which is instead being retained within the network to handle computing tasks. Furthermore, the authors conclude that consolidation is particularly beneficial when the network load is low, while load balancing is to be preferred in high load conditions. Overall, utilizing energy-aware, fast response, and low complexity schedulers results in an increased equivalent computation capacity for the edge network as a whole for a given energy consumption level.

The elastic and predictive online scheduling algorithm for energy harvesting IoT edge networks of~\cite{9403911} demonstrates the achievable benefits that are attainable through careful energy- and RERs-aware redistribution of the processing load. Their Model Predictive Control (MPC) based scheduler solves a sequence of low complexity convex problems, taking as input estimates for future processing and renewable energy arrivals and, based on these, outputs scheduling decisions using a look-ahead approach. This method outperforms a recently proposed scheduling algorithm based on Markov Decision Processes (MDP)~\cite{elasticref} in terms of drop rate and residual energy reserve at the MEC nodes. Moreover, its performance is $90$\% of that of the globally optimal scheduler,  obtained by an offline solver with perfect knowledge about future jobs and energy arrivals.
Along the same lines, \cite{RiskES} investigates energy scheduling solutions to maximize the green energy utilization on microgrid-connected MEC networks. Green energy availability and load demand arrivals are estimated. The optimization objective amounts to minimizing the residual of expected energy and demand by capturing the tail-risk of the uncertain energy and demand arrivals. This risk is used as the input metric for a distributed reinforcement learning mechanism to control the distributed computing resources. The proposed scheme is model-free and learns computing policies in an online fashion, and achieves up to $96\%$ accurate energy scheduling.

\subsection{Mobility Aware Schedulers}
\label{sec:mobility_aware_scheduling}

\paragraph{Vehicular Edge Computing Networks} In the presence of user mobility, computing processes may be migrated among servers for a better user experience. The general idea is that if a mobile user generates a task prior to performing a handover to a different BS, the task may be migrated to the future serving BS, be executed there, and delivered by the new serving BS to the end user once it has moved to the new radio cell. In making these decisions, several considerations arise. If the current server (before the handover) chooses to execute the user task without offloading it, it may degrade the user’s experience. If it decides to migrate the task, it must first choose {\it where} to offload it. Also, it must migrate the corresponding container/VM, so that the destination server can continue the computation without interruption. In addition, deciding where to migrate the task further complicates the resource scheduling problem, as user mobility patterns also come into play. The authors of~\cite{Francesca} devised an online algorithm to estimate the next cell hosting the mobile/vehicle user in a $5G$ vehicular network $4$ seconds before the handover execution, with an accuracy of $88$\%. This method combines Markov Chain (MC) and Recurrent Neural Network (RNN) based predictors, using MCs to forecast the mobility across BS sites and RNN for the fine-grained mobility tracking inside each BS site. Mobility estimation algorithms of this type are key to devising predictive schedulers to proactively allocate resources where they are actually needed.

In~\cite{perin2021ease}, a task scheduling algorithm for \mbox{EH-MEC} networks serving vehicular users is proposed. The objective is to minimize the network carbon footprint subject to task latency and mobility constraints. A centralized \mbox{MPC-based} algorithm is formulated as a first step to predict available green energy and computing resources based on workload arrivals at the BSs and handover probabilities in upcoming time slots. The proposed scheduler prioritizes executing those tasks with high computation intensities and near to expire deadlines, exploiting the available green energy. As a result, it prevents high task migration costs and packet loss. In a second phase, a distributed heuristic determines the best target servers for task migration, utilizing the trajectory predictor from~\cite{Francesca}. As the objective is to minimize the energy consumption from the power grid, servers accept incoming tasks only if the locally available green energy is deemed sufficient to process them in future time slots locally.

\paragraph{UAV-assisted Edge Computing Networks} Infrastructure-assisted UAV systems are a remarkable example of mobility-aware edge networks. There, UAVs may be assigned to execute tasks in predefined positions such as monitoring/surveillance, network coverage extension, data traffic relay, and network assistance in high demand hours. Since UAVs have limited processing power, computing the desired tasks onboard may take longer than with standard network servers, and this may result in excessive energy consumption. To overcome this, UAVs may offload their tasks to high processing power edge servers: this reduces the decision-making time and saves onboard energy. But UAVs have to choose an edge server for offloading their tasks carefully. As the edge server fails/delays the output delivery, they have to consume energy to hover unnecessarily. Authors of~\cite{LevoratoUAV} proposed an optimization decision process where the UAV reads the network state and estimates the task completion time. It then solves an optimization problem to minimize a weighted sum of delay and energy consumption and decides whether to offload the task or not.

\subsection{Joint Resource Schedulers}
\label{sec:joint_optimization}

Network resources (storage, communication, computing, etc.) interact with one another, and their optimization is strongly interdependent. For instance, when a user requests the computation of a task from the edge network, the edge servers may have to communicate among themselves or with a central unit to decide the best suitable server to take the request, and, in some cases, they may benefit from previously cached results. So besides computing, also communication and caching strategies contribute to the energy consumption of the network as a whole. For this reason, many proposed algorithms allocate all network resources {\it jointly}.

\paragraph{Caching and task popularity} In~\cite{Japanese}, a multiuser MEC/BS system with caching capabilities is investigated. End users generate \mbox{computation-intensive} and \mbox{latency-sensitive} tasks (i.e., with hard deadlines). Some tasks are {\it shared} among users, i.e., the same computation outcome may be reused by different users within the same BS coverage area. A {\it task popularity} concept is introduced to mathematically describe the number of users that generate the same task. The general idea is that, in the case of a popular task request, the MEC server selects the connected user that (i) is interested in the task and (ii) has a good (possibly the best) channel quality within the radio cell. This user will upload the request on behalf of all interested users for its task. Hence, the server computes the response for the task and sends it back to the interested users via multicast communications. Again, this saves energy with respect to unicasting the result of the computation to each user. An energy minimization algorithm subjected to deadline and cache size constraints is formulated to jointly allocate {\it memory resources} at the MEC servers and {\it channel access time slots} at the channel access layer. This leads to an \mbox{NP-hard} mixed \mbox{discrete-continuous} optimization problem that can hardly be solved directly and efficiently. Hence, its dual problem (a knapsack problem) and multiple convex problems are respectively devised to allocate memory and channel access resources (time slots). The authors prove that strong duality holds and, thanks to this, obtain the optimal solution for the original formulation. To reduce the algorithm complexity, a suboptimal solution with close-to-optimal performance is also proposed and compared against four reference methods. Two of them allocate equal time to uplink and downlink transmissions, with and without caching. The other two allocate the number of channel access slots proportionally to the size of input and computation result for uplink and downlink transmissions, with and without caching. The proposed suboptimal method outperforms all these reference methods in terms of energy consumption.

\paragraph{Sleep modes} The work in~\cite{DISCO} jointly considers the optimization of access points, edge servers, and user equipment for a computation offloading scenario (from end users to edge servers). Uplink and downlink transmission power, modulation and coding scheme selection, edge server CPU frequency, and duty cycles of all network elements are {\it jointly optimized} towards reducing the total energy consumption. Moreover, a low power sleep mode option is considered for the user equipment, access points, and edge servers. For this setup, a dynamic computation offloading strategy is proposed to minimize the weighted sum of user equipment, access points, and edge servers \mbox{long-term} energy consumption, subject to delay and reliability constraints. By tuning the weights assigned to each energy component, one can switch the focus of the algorithm to \mbox{user-centric}, access \mbox{point-centric}, or edge \mbox{server-centric} optimization. 
Simulation results demonstrate a correlation between access points and user equipment energy consumption. This is because both need to remain active to communicate.  In fact, if the user equipment goes to sleep mode, the access point also has no communication load and switches off. Access points and user centric approach consume more energy than exploiting a \mbox{server-centric} approach. With equal weights, the algorithm reaches \mbox{close-to-optimal} energy consumption for all components. This mode could be useful for utilizing green energy at the maximum possible rate with EH devices at the network and user sides.

\paragraph{\mbox{Non-Orthogonal} Multiple Access (NOMA)} The authors of~\cite{NOMA1} study the impact of \mbox{NOMA-based} scheduling on the outage probability of \mbox{energy-harvesting} IoT devices for a {\it cooperate and transmit scenario}. Their analysis shows that leveraging NOMA for communication can decrease the outage probability by almost $10$\% with respect to TDMA and signal combining methods under a Rician fading channel. Further, in~\cite{NOMACHINA} a \mbox{NOMA-based} \mbox{energy-efficient} task scheduling algorithm utilizing a Self-Imitation Learning (SIL)-based DRL is proposed to capture the \mbox{time-varying} movement and load pattern in a {\it vehicular edge computing} scenario. The proposed algorithm jointly optimizes downlink transmission/computing powers and task transmissions among RSUs, by meeting service delay requirements. The proposed method is compared against (i) a greedy algorithm that assigns offloaded tasks to the MEC server with the minimum required energy consumption to satisfy the latency constraint, (ii) a reference scheduling algorithm where edge servers accept all the tasks offloaded by the passing vehicles without balancing the load among them. Simulation results demonstrate that, thanks to NOMA transmissions, the energy consumption decreases with an increasing number of subchannels and that the number of completed tasks also increases. Performance improvements gradually decrease, becoming negligible as the number of subchannels increases beyond $5$. The proposed solution outperforms the other methods, serving more users across all ranges of vehicular speed for a lower or comparable energy consumption.

\paragraph{Device to Device (D2D) communications and computing} D2D transmissions can help MEC networks achieve \mbox{low-latency} service goals. Thanks to D2D communications, third-party fog nodes may transmit data to nearby nodes by reusing their channel bandwidth resources without having to route this data through an access point. Network operators may rent third-party nodes to use as data relays and sometimes even as computing resources. As a result, MEC networks can leverage D2D communication {\it to increase spectral efficiency} and data rate, {\it fulfill low-latency requirements}, and {\it avoid excessive energy consumption}. However, reusing non-orthogonal frequency resources in the channel bandwidth causes inter-cell interference, pilot contamination, and \mbox{signal-to-noise} ratio challenges in estimating channel state information and signal detection~\cite{PASANGI}. Thus, in \mbox{D2D-assisted} MEC networks, it is strategic to consider the allocation of communication resources such as channel bandwidth, data rate, interference, and transmission power along with computation resources. The authors of~\cite{downlinkD2D} exploit D2D communications to relieve the load on edge servers through video content {\it caching} and downlink transmissions. They propose two resource allocation algorithms to maximize the {\it energy efficiency} of the proposed D2D transmission system through the optimization of channel access resources and transmission power, with constraints on downlink interference and data rate. Users can simultaneously retrieve their required data from edge servers or nearby devices through a hybrid caching and transmission scheme. Since the formulated energy efficiency maximization problem is NP-hard, it is transformed into a mixed-integer nonlinear program using Dinkelbach's method to facilitate the search for a solution. Hybrid transmission schemes are considered to avoid co-channel interference, leading to higher throughput and energy efficiency than the four reference methods. Nevertheless, with an increasing number of users, the strong \mbox{co-channel} interference makes it difficult to maintain high energy efficiency, posing additional challenges to the proposed solution. It is noted that this work does not involve computation at the end nodes but rather uses them within an edge caching system to increase the reliability and efficiency associated with the relay of video content to end users.
 
\paragraph{Reconfigurable Intelligent Surfaces (RISs)} The authors of~\cite{Greenedgeinf} propose an optimization technique to improve the energy efficiency of an edge inference system with RISs. In their network setup, \mbox{resource-constrained} mobile devices generate inference tasks and offload them to \mbox{resource-constrained} edge servers. An energy consumption minimization problem is formulated by jointly addressing the task selection strategy for the edge servers, transmit/receive beamformer vectors at each BS, transmit power, and the RIS \mbox{phase-shift} matrices for downlink/uplink transmissions. This leads to a combinatorial problem with coupled optimization variables. To devise a low complexity solver, the sparse nature of beamforming vectors is leveraged, along with their interaction with other problem variables. In particular, the group sparsity nature of variables makes it possible to obtain a \mbox{three-stage} solution, splitting the original problem into simpler subproblems. Simulation results demonstrate the effectiveness of the proposed technique, which reduces the overall network energy consumption by $25$\% in the low Signal-to-Interference-plus-Noise Ratio (SINR) regime and by $45$\% in the high SINR regime.

\subsection{Remarks on resource allocation}
\label{sec:resource_allocation_open_challenges}

While hardware characteristics certainly have a substantial impact on system performance, no less critical are the algorithms that are employed to utilize that hardware, since ill-designed management algorithms may easily nullify any effort made to improve the energy efficiency and environmental sustainability of the system. Therefore, the strategies determining how to allocate RERs, computing,  communication, and storage resources among the users are certainly of fundamental importance. Among the key aspects to consider for an effective scheduling policy, there are:
\bi 
\item the {\it prediction} of exogenous processes, e.g., the energy income from RERs and the processing load from the UEs,  
\item  the objective function that is used to define and solve the resource allocation problem. We underline that the objective, in this case, differs from the sole optimization of user/network Quality of Service (QoS) metrics such as delay and data transfer rates (throughput). In fact, in the presence of RERs, it is strategic to use the energy when and where it is available from renewable sources, e.g., when there is an {\it energy surplus}, as otherwise the excess energy may be lost, especially if the MEC nodes do not have energy storage capability (batteries).
\item the design of {\it low complexity} solvers to obtain the optimized schedule, i.e., to allocate and possibly redistribute the load across the MEC servers.
\ei
We summarized the novelty aspect of proposed resource scheduling algorithms in table~\ref{tab:resource_allocation}.

\begin{table*}[tb]
\footnotesize
\begin{center}
\begin{tabularx}{0.92\textwidth}{Y|c|c|c}
\toprule
\textbf{Papers}&\textbf{Objective}&\textbf{Metric} & \textbf{Impact} \\
\midrule
~\cite{9197634}& \makecell{Joint network and task\\offloading ratio optimization} & \makecell{Energy, CPU speeds,\\bandwidth, transmission power} & \makecell{Minimizing computational burden on MEC\\ network with high energy efficiency}\\
\midrule
\cite{9380662}& \makecell{Maximizing completed tasks\\with latency/energy constraints } & \makecell{Energy, Delay} & \makecell{Increased energy efficiency and\\completed task count}\\
\midrule
\cite{9054646}& \makecell{Maximizing RER utilization} & \makecell{Time, state of the RER,\\Battery, load and BS} &  \makecell{ High energy efficiency with\\reduced carbon footprint }\\
\midrule
\cite{perin2022towards, 9403911}& \makecell{Load balancing/consolidation,\\Energy management,} & \makecell{Energy, Latency} &  \makecell{Increased energy efficiency and\\computation capacity per energy unit}\\
\midrule
\cite{RiskES}& \makecell{Minimizing the RER residual} & \makecell{Energy, Load, and Network state} &  \makecell{Accurate and efficient RER scheduling}\\
\midrule
~\cite{perin2021ease}& \makecell{Carbon footprint minimization} & \makecell{Energy, Latency} & \makecell{Minimizing carbon footprint \\ within QoS constraints}\\
\midrule
\cite{LevoratoUAV}& \makecell{Optimizing offloading decisions} & \makecell{Energy, Latency} & \makecell{Increased energy efficiency \\ and task completion count}\\
\midrule
~\cite{Japanese}& \makecell{Caching and task offloading} & \makecell{Energy, Latency} & \makecell{Increased energy efficiency}\\
\midrule
\cite{DISCO}& \makecell{Radio/computing resource scheduling} & \makecell{Energy, Delay, Error rate} & \makecell{Close-to-optimal sub-optimal solution}\\
\midrule
\cite{NOMA1, NOMACHINA}& \makecell{Energy efficient task scheduling} & \makecell{Energy, Latency} &  \makecell{Reduced energy consumption}\\
\midrule
\cite{downlinkD2D}& \makecell{Caching and transmission scheme}& \makecell{Energy, Data rate, Bandwidth} &  \makecell{Increased energy and spectral efficiency}\\
\midrule
\cite{Greenedgeinf}& \makecell{Energy efficient task scheduling} & \makecell{Energy, QoS} &  \makecell{Reduced total energy consumption}\\
\bottomrule
\end{tabularx}
\end{center}
\caption{Impact of the proposed energy efficient resource schedulers from the literature on energy consumption.}
\label{tab:resource_allocation}
\end{table*}

\section{Energy Efficient Collaborative Learning at the Network Edge}
\label{sec:collaborative_learning}

Massive datasets and spare computational power are abundant at the network edge. Collaborative edge learning algorithms allow leveraging edge computational powers to train ML models locally, using the data collected by end nodes. Specifically, learning distributedly a model over $N$ edge devices with global parameters $\bm w$ and dataset $\mathcal{D}$ amounts to solving the problem
\begin{equation}
    \label{eq:distributed_learning}
    \min \; \left[F\left(\bm w \mid \mathcal{D}\right)=\sum_{i=1}^N p_i F_i(\bm w_i \mid \mathcal{D}_i)\right],
\end{equation}
for a set of weights $p_i$, where $F_i$ are the local {\it loss functions} and $\bm w_i$ and $\mathcal{D}_i$ are the local model parameters and datasets, respectively.

\begin{figure}[tb]
\centering
\includegraphics[width=\scaledwidth{0.55}]{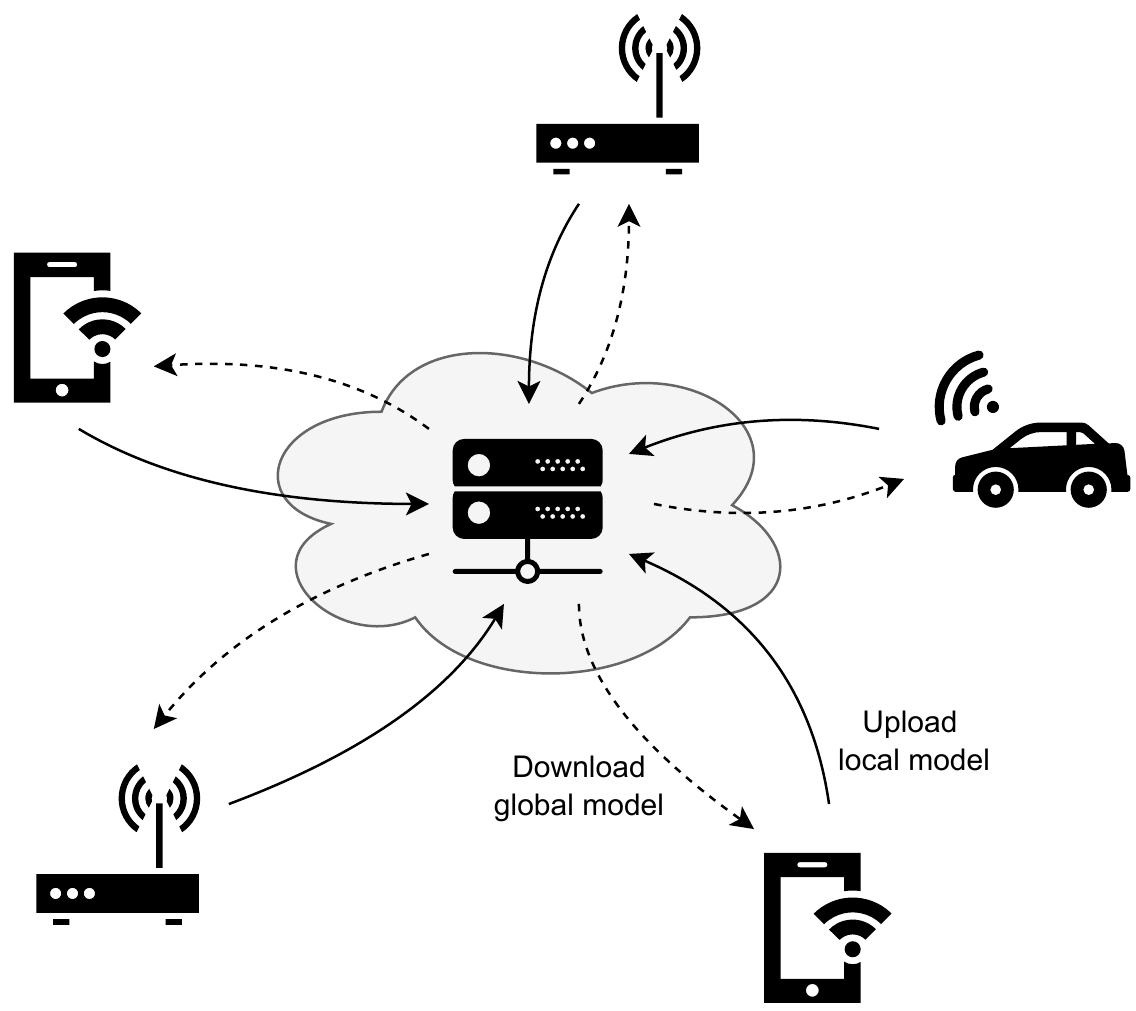}
\caption{\textbf{Federated learning.} The edge devices, acting as clients, locally train the model with their own data and periodically upload it to the server. The server aggregates the local models coming from the clients and send to them an updated version of the global model.}
\label{fig:federated}
\end{figure}

\begin{figure}[t]
\centering
\includegraphics[width=\scaledwidth{0.55}]{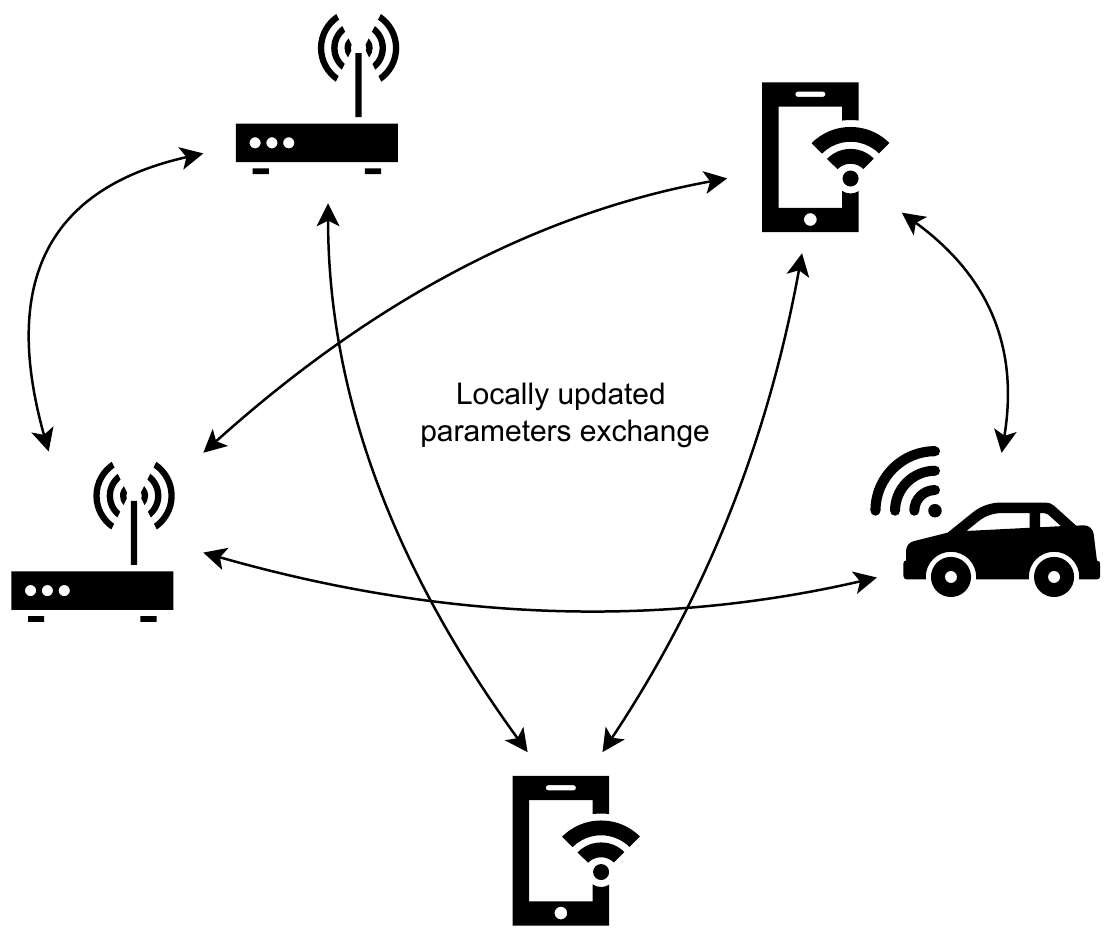}
        \caption{\textbf{Fully decentralized learning.} Edge devices are connected via a general mesh topology. They update the model parameters using some message passing algorithm, by periodically communicating with their neighbor nodes and without involving any central aggregator.}
    \label{fig:decentralized}
\end{figure}

Distributed ML models are appealing due to privacy considerations, as the local data that is collected at the distributed node sites does not have to be disclosed (sent). Local computations are instead performed at the nodes using their own data, and only model updates (gradients) are sent across the network. Nevertheless, although this is highly attractive, it also presents challenges related to the time-varying nature of computational power, communication rate, and energy availability at each device. Furthermore, communication, energy, and computation resources are often highly heterogeneous across wireless and edge devices, which leads to further complications. Hereafter, we refer to the difference in communication/computing/energy resources across nodes as \emph{system heterogeneity}, while the fact that the data may be non-i.i.d. across devices is referred to as \emph{statistical heterogeneity}. Utilizing heterogeneous edge resources to accurately train ML models in an energy-efficient manner is a challenging and still open endeavor. Distributed Learning (DL) algorithms come in two primary flavors: (i) \emph{Federated Learning} (FL)~\cite{mcmahan2017communication}, where the workers are connected with a star topology to a physical unit serving as an aggregator for their local models (see Fig.~\ref{fig:federated}), or (ii) methods obtaining the ML model as the result of a \emph{fully decentralized} communication/learning process, where workers share their locally updated ML model with their direct neighbors in any generic mesh topology (Fig.~\ref{fig:decentralized}), progressively reaching consensus on a common model.

\subsection{Context-Agnostic Efficient Communication}
\label{sec:FED_context_agnostic}

The amount of data exchanged for the model updates represents one of the major DL challenges (the communication bottleneck problem) due to the large size of typical ML models. This affects both the convergence time and the energy drained to transmit the model updates. According to~\cite{D.Gunduz+V.Poor}, context-agnostic approaches to tackle this problem subdivide into (i) \emph{sparsification}, i.e., pruning of some elements of the local gradients, (ii) \emph{quantization}, i.e., reducing the number of bits to encode the local gradients, and (iii) \emph{local Stochastic Gradient Descent (SGD)}, namely, performing multiple local gradient descent steps before performing the aggregation. The first two methods reduce the number of bits sent for the aggregation steps, while the latter aims to reduce the total number of aggregation rounds needed.

\subsection{Context-Aware DL Optimization}
\label{sec:FED_context_aware_DL_opt}

Despite the fact that the mechanisms mentioned above can be combined, this is not enough to make communication efficient in large-scale DL setups. In fact, the above strategies are independent of the communication channel and protocols used. A higher level of context awareness (in this case, channel characteristics and communication resources) in wireless networks holds the potential of significantly improving the efficiency of model training, as the channel quality varies across network devices and over time, and local resources such as bandwidth and transmission power are limited. Scheduling and resource allocation thus assume a prominent role in improving the efficiency of DL.

\paragraph{Client selection and resources allocation} \emph{Client selection} refers to picking, at each learning step, a subset of clients that contribute to updating the model parameters within the global aggregation phase, hence saving network resources. As a simple approach, scheduling algorithms could select those clients with the best experienced channel. This choice, although speeding up the model training by avoiding the \mbox{so-called} \emph{stragglers}, may result in a biased ML model in the presence of statistical heterogeneity, polarizing it towards the local models of those clients with a better channel quality, e.g., better positioned with respect to the access point. To cope with this, the authors of~\cite{D.Gunduz+V.Poor} propose a scheduling policy that considers the age of information (AoI) (i.e., the number of iterations from the last update) of each client in conjunction with the experienced channel quality. By penalizing this metric, they obtain a better performing model by using network (communication) resources efficiently. The use of analog transmissions to aggregate the model over-the-air is proposed in~\cite{sun2022dynamic}, where client selection is performed at the beginning of each round by optimizing (with a Lyapunov approach) a learning bound obtained in the paper, subject to communication and processing energy constraints. This procedure is shown to perform close to an optimal benchmark in terms of accuracy while outperforming a myopic scheduler. In papers~\cite{tran2019federated, yang2021federated, xiao2021vehicle} a convergence result is exploited to bound the squared norm of the gradients of the local model loss functions as
\begin{equation}
    \label{eq:learning_bound}
    \left\Vert \nabla F_i\left(\bm{w}_i^k\mid\mathcal{D}_i\right)\right\Vert^2\le\theta\left\Vert \nabla F_i\left(\bm{w}_i^{k-1}\mid\mathcal{D}_i\right)\right\Vert^2,
\end{equation}
for some $0 \le \theta \le 1$. In particular, when $\theta=0$, the model will get an exact solution to the local problem, while if $\theta=1$, no progress is observed between iterations $k-1$ and $k$. If the objective is strongly convex, the convergence rate is proven to be $\mathcal{O}\left(\log\left(1/\theta\right)\right)$.
In~\cite{tran2019federated}, the joint optimization of a weighted average of computing, transmission energy, and training time is considered. In their scenario, the server can select the proper bandwidth to communicate with the clients to adjust transmission energy and time while the clients decide their processing frequency locally. The authors of~\cite{yang2021federated} realized a customized frequency division multiple access (FDMA) strategy to jointly optimize the energy used for communication and local computations by selecting the CPU frequency, the transmission power, and bandwidth and also providing a formal convergence analysis of the proposed algorithm. Their scheme outperforms selected benchmarks, including a time division multiple access (TDMA) approach, when looking at the tradeoff between energy and completion time. Further, in~\cite{xiao2021vehicle} a vehicular scenario is considered, where the authors perform both vehicle selection and resources allocation, again optimizing completion time and energy. The selection task is performed in a greedy manner to get higher image quality data, while the resource allocation is formulated as a min-max problem, devoted to minimizing the time and the energy needed for the worst scheduled vehicle. Due to the absence of benchmark algorithms for a vehicular environment, the authors show a comparison with the performance obtained by optimizing the accuracy only or the communication resources only, demonstrating that their framework leads to fairer solutions.

\paragraph{System heterogeneity} Heterogeneous capacities of edge devices is a crucial problem. Some may be equipped with a GPU, while others only have the CPU. Some may be interconnected via unreliable wireless channels, whereas others through optical fiber backhaul links. System heterogeneity is the special focus of papers~\cite{wang2019adaptive, luo2021cost, li2021talkorwork, kim2021autofl}. The authors of~\cite{wang2019adaptive} at first theoretically analyze the convergence rate of distributed gradient descent algorithms and then develop an adaptive control framework to find the best tradeoff between the number of local steps at the distributed nodes (the clients) and the number of model aggregation rounds, optimizing theoretical learning bound subjected to a local energy budget. The proposed network controller is evaluated on both convex problems and neural networks, obtaining close to optimal performance. Of particular note, this also holds for models based on Support Vector Machines (SVMs). The same objective is investigated in~\cite{luo2021cost}, where client selection is also addressed besides the number of local steps for the clients. The problem is formulated as multi-objective optimization, considering a weighted energy and completion time average, subject to an expected accuracy constraint. A real testbed implementation comprising $30$ devices with heterogeneous computing capacities has been carried out, and results for neural networks training show that the proposed scheme is faster to converge and globally consumes less energy against competing solutions while also achieving a better model accuracy. The approach suggested in~\cite{li2021talkorwork} additionally involves personalization of the gradient sparsification strategy for each client, where the (gradient) compression parameters are decided by evaluating the available energy resources for computation and communication. The proposed scheme drains significantly less energy with respect to benchmark algorithms in the presence of highly heterogeneous systems, almost without affecting the test error. An RL framework for client selection called AutoFL is proposed in~\cite{kim2021autofl} to obtain a data-driven solution to the problem of system heterogeneity, with the objective of optimizing training time and energy efficiency. Taking as input the resources that are locally available to each client, AutoFL obtains a speed gain of $3.6$ and an energy efficiency gain of about $5$ times over traditional methods.

\paragraph{Fully decentralized learning} The research described in the previous paragraphs considers federated learning solutions, where a central aggregator is in charge of updating the model from inputs received from a set of clients. The world of {\it fully decentralized learning} is instead widely unexplored when it comes to energy optimization. Although context-agnostic methods can always be applied to improve the energy efficiency and the utilization rate of resources, an essential and unique feature of decentralized learning is the possibility of adjusting the communication network topology. In~\cite{kuo2021energy}, the authors select a subset of the available links to perform model broadcasting so as to minimize the transmission power levels. A constraint on the minimum number of links required to guarantee the convergence of learning is added. This results in a combinatorial NP-hard problem over a graph that is relaxed to obtain an approximate solution. The considered scenarios involve the presence of interference and packet collisions. The results show that the proposed optimization can reduce the energy consumption by more than $20\%$ compared to simple heuristics, without impacting the model performance.

\begin{table*}[t]
\footnotesize
\begin{center}
\begin{tabularx}{0.92\textwidth}{Y|c|c|c}
\toprule
\textbf{Papers}&\textbf{Objective}&\textbf{Metric} & \textbf{Impact} \\

\midrule
\cite{D.Gunduz+V.Poor}& client selection & channel quality \& AoI & reducing model bias  \& energy efficiency\\
\midrule
\cite{sun2022dynamic}& over-the-air client selection & learning bound w{/} energy constraints & good model accuracy\\
\midrule
\cite{tran2019federated, yang2021federated, xiao2021vehicle}& resource allocation & \makecell{CPU, TX power \& time\\ w/ target accuracy} &  \makecell{fair  network resource allocation\\ \& time/energy trade-off}\\
\midrule
\cite{wang2019adaptive}&CPU/TX trade-off& learning bound  w/ energy constraints& \makecell{close-to-optimal performance \\ in heterogeneous systems}\\
\midrule
\cite{luo2021cost,kim2021autofl}&\makecell{CPU/TX trade-off \\ \& client selection}& \makecell{time/energy trade-off \\ w/ target accuracy}& reducing completion time  \& global energy\\
\midrule
\cite{li2021talkorwork}&gradient sparsification&TX power& reducing energy without affecting accuracy\\
\midrule
\cite{kuo2021energy}&\makecell{topology selection \\ (fully decentralized)}&TX power& reducing energy without affecting accuracy\\

\bottomrule
\end{tabularx}
\end{center}
\caption{Energy efficiency impact of scheduling algorithms in collaborative learning.}
\label{tab:impact_of_fedlearn}
\end{table*}

\subsection{Remarks on collaborative learning}
\label{sec:FED_open_challenges} 

The decentralized computing facilities of the edge platform can be effectively exploited to train ML models. A key aspect concerning this aspect is the management of the sometimes highly different devices to avoid wasting energy and time while meeting the desired model accuracy. The main take-home messages of this section are summarized below.
\begin{itemize}
    \item Effective approaches tackling the device/client selection and resources allocation problems can either optimize network and computing resources while considering a learning bound constraint or a soft penalty on some fairness metric, e.g., the AoI or do the dual, i.e., minimizing a theoretical learning bound with some resources budget.
    \item System heterogeneity is a relevant issue in DL, and literature is currently focusing on finding solutions to the problem of the computation-communication tradeoff, i.e., deciding how many local SGD steps each device should perform before aggregating the model.
    \item When fully decentralized learning is used instead of FL, another way to tune the energy used and the convergence time is to adjust the virtual network topology.
\end{itemize}

A summary of the papers presented in this section, their technical proposal, and their main impact is reported in Table~\ref{tab:impact_of_fedlearn}.

\section{Discussion and Open Challenges}
\label{sec:conclusions}

Mobile networks are getting energy hungry, and this trend expects to continue due to a surge in communication and computation demand. MEC networks, a key component of $5G$ and $6G$ networks, will entail energy-consuming services and applications. It is critical to make them as energy efficient as possible, exploiting renewable energy resources, new algorithmic designs, intelligent provisioning and resource allocation strategies, etc. In the following, we identify open research avenues and challenges for each of the aspects covered within the present survey, underlying the importance of precise and up-to-date servers' energy consumption models.

\paragraph{MEC energy profiling}
Some works have recently appeared on EH-MEC networks and their integration with the power grid. However, there are missing aspects to be investigated and integrated into EH-MEC study cases to increase their validity, as follows. We advocate that most of the energy consumption models mentioned in section \ref{sec:Emodel} are no longer adequate to quantify the energy consumption of modern CPU/GPU edge processors and, in turn, to devise effective resource schedulers. Some initial works have recently appeared, e.g., covering NVIDIA Jetson TX~\cite{Jetson-profiling-2020-1} and Nano boards~\cite{Jetson-profiling-2020-2}. However, these models are neither complete nor sufficient to obtain appropriate objective/cost functions that accurately reflect the energy consumption of edge servers within a resource scheduling problem. Aspects that should be better investigated and reflected into proper cost functions are: 
\begin{enumerate} 
\item the timing behavior and energy drained by turning on and off containers or virtual machines~\cite{Tao-2019},
\item the energy consumption of the servers on the idle state and in general on each of the available power saving modes, as well as the time required to put them into such states or wake them back up when idling,
\item accurate models to gauge the energy expenditure of ML processes running sequentially, in parallel, or on overlapping time frames~\cite{Aria&Ahmed}.\\
\end{enumerate}
Furthermore, in our review of the initial design step of MEC networks, i.e., offline MEC Network Deployment optimization methods, we noticed that energy efficiency and sustainable design parameters, e.g., carbon footprint and RER availability, have been widely neglected. A discussion on this follows.

\paragraph{MEC deployment} 
Compared to the extensive body of literature looking at the user/network QoS, only a few works analyze {\it energy efficient} MEC deployment strategies. Also, since MEC deployment strategies are offline, researchers may utilize more complex optimization tools than, e.g., the simple heuristics in~\cite{mobilityawareMECdeployment}, to determine MEC deployment strategies based on a complete set of Key Performance Indicators (KPIs), such as outage, load balance, energy efficiency, and in particular {\it expected energy harvesting figures}. We remark that existing designs neglect the latter in the literature. Furthermore, leveraging time variable load patterns and the possibility of using Autonomous Unmanned Vehicles (UAVs) to selectively provide MEC support in areas with a temporarily high load may introduce a higher degree of flexibility and reduce the number of servers that are to be deployed~\cite{joinRA&UAVplacement}.

As for resource and service provisioning algorithms, we analyzed their inherent difference with dynamic resource allocation algorithms, pointing out their impact on energy efficiency and system sustainability. In the following, we discuss the related open challenges and possible future research directions.

\paragraph{Reference computing models} The computing tasks are often abstracted and identified in terms of quantities such as data size and computation intensity, e.g., the number of CPU cycles required per input byte. The actual connection of such quantities with real world computing processes is often ignored or modeled in a simplistic manner. To our understanding, this is due to the lack of data sets and characterization studies covering modern computing tools and needs for mobile users and IoT systems. Often, papers indicate future scenarios such as virtual reality and autonomous driving, but specific computing architectures, their energy consumption models, and realistic workload arrival statistics are rarely found. Reference computing models would be very beneficial to the design and the quantitative evaluation of resource provisioning algorithms. 

\paragraph{Caching and memory management} The energy consumption drained by content caching at the edge servers were not considered in previous research. This follows the rationale that cached data is reused several times, so the energy cost of storing it becomes negligible concerning the other operations carried out (on or with it) at running time. However, this may change according to the ratio between the number of caching actions per time interval and the number of times that the cached data is reused during the same time interval. Keeping this into account may lead to different considerations for the caching policy. Likewise, the impact of memory read/write operations on the energy consumption at the edge nodes, e.g., for storing a model prior to its use, is often ignored and deserves further attention.

\paragraph{Hardware aging} Service providers tend to implement 5G infrastructure upon the previously installed network entities. Electronic devices' energy efficiency decreases over time. The impact of the aging of edge computing devices is a totally unexplored area and maybe also a relevant aspect to look at.

\paragraph{D2D aided computing and transmissions} Most D2D-aided MEC resource allocation techniques in the literature consider spectrum reuse for uplink transmissions, as spectrum reuse in the downlink channel suffers from inter-cell interference due to high power signals transmitted by the BSs~\cite{D2DOP}. We advocate that D2D communications could be used in conjunction with intelligent transmission power control schemes to alleviate interference problems when supporting edge traffic. So, the combined use of D2D transmissions, power control (downlink), and fog computing holds the potential to increase the energy efficiency of BS/MEC networks. Moreover, in addition to using D2D strategies for communications, there is room for also using them to carry out some of the computation or for intelligently processing user traffic (see, e.g., the following paragraph). 

\paragraph{Semantic compression} Transferring unfiltered data from end IoT devices to the edge servers through the (often) limited available bandwidth of communication channels might cause packet loss and reduce the bandwidth available to other services. The authors of~\cite{Levoratosemantics} propose a semantic-based data selection and compression approach for IoT sensors. In their proposed method, low complexity linear classifiers are distributed to end IoT devices. Such classifiers are aligned with a global non-linear classifier located at the edge server, having high-level operation objectives. The classifiers at the IoT nodes filter their generated data stream to alleviate the communication burden and the bandwidth usage over the wireless links.
On the other hand, the edge server continuously updates the local classifiers at the end nodes based on its analysis of the received data. As expected, a tradeoff exists between the energy consumption of IoT sensors,  their bandwidth usage, and the classification quality at the edge server. Also, the amount of transferred data is bounded by the time-varying bandwidth capacity, the available energy, and the processing power at the IoT nodes. The reference \cite{Levoratosemantics} is the only one we found on this research line, which remains largely unexplored. We believe that semantic processing/compression is a sensible and promising approach. Understanding the most extensive compression level to be used at the IoT nodes so that the main server can still make correct classifications/decisions is highly signal- and application-dependent. Further research is required to analyze semantic compression approaches as a function of energy/bandwidth resources/constraints, considering real world applications.

Lastly, we emphasized the importance of energy efficiency for collaborative learning at the network edge. These algorithms are expected to be highly and consistently utilized, so their energy efficient design is a must. Possible research directions are as follows.

\paragraph{Distributed learning approaches} Since methods for sparsification and quantization are well established in information theory, current research focuses on strategies to better exploit local SGD. Specifically, the problem of client selection and resource allocation is being investigated also in the presence of system heterogeneity, but (i)~current work lacks the integration of future resource availability predictors. This is particularly important when dealing with vehicular scenarios, where trajectory forecasting, along with a spatial characterization of the environment, would lead to an improved context awareness (and to a consequent increase in the attained performance).
The trade off between the number of local steps and aggregation rounds needs to be better addressed by (ii)~exploiting federated algorithms that are specifically designed for heterogeneous systems~\cite{wang2020tackling} and by (iii)~using learning bounds that better adapt to problems for which strong convexity does not hold, such as neural networks. Finally, (iv) current research mostly focuses on FL, while energy optimization for decentralized learning is still unexplored, despite this paradigm being extremely attractive for cross-silo and peer-to-peer collaborative learning.

\begin{acks}
The present work has received funding from the European Union’s Horizon 2020 Marie Sk{\l}odowska Curie Innovative Training Network, GREENEDGE (GA. No. 953775), and by the Italian Ministry of Education, University and Research (MIUR) through the initiative ``Departments of Excellence" (Law 232/2016). The views and opinions expressed in this work are those of the authors and do not necessarily reflect those of the funding institutions.
\end{acks}

\bibliographystyle{ACMReferenceFormat}
\bibliography{main}

\end{document}